\documentclass[aps,pre,preprint,groupedaddress]{revtex4-1}
\usepackage{graphicx}
\usepackage{amsmath}
\usepackage{float}
\usepackage{verbatim}
\usepackage{ amssymb }
\usepackage{graphicx}
\usepackage{color}

\begin{document}
\title{Extinction Dynamics of Cardiac Fibrillation}
\author{David Vidmar}
\affiliation{Department of Physics, University of California, San Diego,
La Jolla, CA 92093}
\author{Wouter-Jan Rappel}
\affiliation{Department  of Physics, University of California, San Diego, La Jolla, CA 92093}

\begin{abstract}
During episodes of atrial fibrillation, the heart's electrical activity becomes disorganized and shows fragmenting spiral waves.
To systematically address how this pattern terminates using spatially extended simulations  exceeds current computational resources. 
To circumvent this limitation, we treat the number of spiral waves as a stochastic population with a corresponding birth-death equation and use techniques from statistical physics to determine the mean episode duration of atrial fibrillation. 
We show that this duration can be computed for arbitrary geometries in minimal computational time and that it depends exponentially on tissue size, consistent with the critical mass hypothesis which states that fibrillation requires a minimal organ size. 
Our approach can result in efficient and accurate predictions of mean episode duration, thus 
creating a potentially important step towards improved therapeutic interventions for atrial fibrillation.
\end{abstract}

\maketitle

\section{Introduction}

Atrial fibrillation (AF) is the most common cardiac arrhythmia, affecting millions of people worldwide  \cite{Chuetal13}.
During AF, the conduction pattern within atrial tissue becomes
irregular, resulting in serious health consequences  including
heart failure, stroke, and mortality \cite{fuster2006acc}.
Although the precise mechanisms for AF are still unclear, 
conduction patterns involving localized spiral waves 
(spiral waves with a tip location that remains within a certain spatial domain) 
have been shown to be a crucial component in the maintenance of this disease \cite{jalife2003rotors, nattel2017demystifying, Naretal12b}.
The stable driving sources of AF can be eliminated using targeted 
ablation which destroys tissue. 
After the removal of driving sources,  spiral wave reentry results in multiple and often 
migratory and transient waves of activation \cite{Naretal12a,haissaguerre2016intermittent, Naretal12c}. These spiral waves continuously break down to form new 
ones, and are removed through collisions with other spiral waves or with 
non-conducting boundaries 
as has been shown by many computational studies 
\cite{karma2013physics,Fenetal02,zykov2017fast,fink2011cardiac,qu1999cardiac}. This stochastic competition between 
creation and annihilation results in   
spiral defect chaos (SDC), a dynamical state  
described in a variety of different excitable systems 
\cite{coullet1989defect,Egoetal00, DanBod02,beta2006defect}.
SDC persists until the last spiral wave is terminated, with its duration representing a stochastic event.
 
Critical in AF management and therapy is thus the mean episode duration $\tau$ which is a statistical measure of the average time of reversal to normal sinus rhythm. 
The ability
to infer $\tau$ from some observables of our system, particularly in the presence of 
different surgically created lesion sets and pharmacological interventions, can be an important step towards more efficient AF therapies and 
patient-specific ablation procedures \cite{rotter2007impact, mcdowell2012methodology}. 
Unfortunately, determining $\tau$ through direct simulations of 
spatially extended cardiac models is challenging because
a statistically significant quantification of this stochastic quantity 
requires the time-consuming task of simulating a multitude of episodes
\cite{virag2002study, Dosetal12}. 
This becomes even more problematic for large geometry sizes since 
$\tau$ increases sharply as a function of the system size  
\cite{Qu06}. This increase is related to the so-called critical mass hypothesis which 
posits that fibrillation requires atria with a minimal size \cite{garrey1914nature, Byretal05}. A fundamental 
elucidation for this hypothesis is currently lacking.

 In this study, we develop an alternative method to compute $\tau$ by treating the
number of spiral tips $n$ as a stochastic quantity and casting 
 its birth-death process into a master equation, 
a commonly used approach in the field of population
dynamics \cite{lande2003stochastic,van1992stochastic,Gard_book09}.
This approach was also used in recent studies which examined  filament turbulence in phenomenological models and which described 
the dynamics of surface defects in terms of a master equation   \cite{davidsen2008filament,reid2011filament,st2015influence}.
Contrary to these studies, we focus here on tips migrating in 2D and on termination events and the associated mean episode duration.
 In our case, the master equation 
 describes  the probability $P(n,t)$ of  having $n$  spiral tips at
  time $t$ as
 \begin{eqnarray}
 \frac{dP(n,t)}{dt}= \sum_r \left[ W_{r}(n-r)P(n-r,t)-W_{r}(n)P(n,t) \right]
 \label{eq_master}
 \end{eqnarray}
 where $W_{r}$ 
 are transition rates for the number of spiral tips to change by $r$ tips and can be computed 
directly from spatially extended simulations of cardiac models.
Since tips are created and annihilated either as pairs or as singlets, we only need to consider $r=\pm1, \pm 2$.
  As a boundary condition we  take $n=0$ to be absorbing. 
  This means that there is no escape from the no-tip state and that all birth rates for $n=0$ vanish: $W_r(0)=0$.
  Furthermore, an additional boundary condition stems from the fact that for $n=1$ the pair-wise death rate equals 0: $W_{-2}(1)=0$.
  Once the  rates are known, we can construct a
  transition matrix which can be used to compute $\tau$ at minimal 
  computational cost \cite{newman2010networks}.

 For large $n$, the death rate will exceed the birth rate since tips will 
 have a high probability of colliding. As a result,  
 the number of spiral tips does not grow to very large numbers. 
 If for small $n$
 the birth rate is larger than the death rate, then a long-lived
 (quasi-stationary) metastable
 state exists with a mean number of tips $\bar{n}$. 
 The distribution associated with this metastable state 
 is called the   quasi-stationary distribution $P_{qs}(n)$ 
 \cite{dykman1994large,assaf2010extinction}. 
 Note that for systems with an absorbing state at $n=0$, the stationary distribution trivially corresponds to $P(n)=0$ for all $n\ne 0$ and $P(0)=1$  
In the quasi-stationary state, the number of tips fluctuates around the average  value for prolonged periods of time and the mean episode duration can be computed using
\begin{eqnarray}
\frac{1}{\tau} = \sum_{r<0} W_r(-r) P_{qs}(-r) .
\end{eqnarray} 
Termination only 
 occurs during rare escape events, corresponding to a large fluctuation away from the mean number of tips. As a consequence, standard equilibrium statistical 
 physics approaches based on small fluctuations do not apply  
 \cite{assaf2010extinction,doering2005extinction}. 
 Instead, techniques 
from non-equilibrium statistical physics must be invoked to determine statistical quantities corresponding to extinction, including $\tau$.

To illustrate our stochastic approach to quantifying termination dynamics,
we carry out simulations of SDC using spatially extended electrophysiological models.
We should stress, however, that the approach should also work for  
other systems that exhibit spiral wave dynamics, including the
complex Ginzburg-Landau equation 
\cite{aranson2002world} or simple phenomenological models \cite{barkley1990spiral}.
 The ``direct'' simulations 
use the standard reaction-diffusion equation:
\begin{eqnarray} 
{\partial_t V}= D\nabla^2 V -I_{ion}/{C_m}
\end{eqnarray}
where $V$ is the transmembrane potential, $C_m$ ($\mu \rm{F}\, cm^{-2}$) is the membrane capacitance  and $D \nabla^2$ expresses the inter-cellular
coupling via gap junctions and diffusion constant $D$. 
The membrane currents in 
the electrophysiological model are denoted by
$I_{ion}$ which are governed by nonlinear evolution equations coupled
to $V$. For our purposes, the precise form of $I_{ion}$  is not important and
we present results using the detailed Luo-Rudy (LR) model \cite{LuoRud91}, modified to obtain spiral wave break-up as described in
Qu {\it et al.} \cite{Quetal00b}. To stress the generality of our approach we have also carried  simulations using the  
simplified  Fenton-Karma (FK) model  (parameter set 8) \cite{FenKar98}. 
Results from these simulations are shown in the 
Supplemental Material \cite{supplemen}. We perform the simulations in  square 
two-dimensional  computational domains although our approach can be  equally well applied in more complex
geometries.  As boundary conditions, we  consider both 
non-conducting and periodic boundary conditions, and we vary the area of the 
computational domain, which is equivalent to varying  $D$ while keeping the area constant. 
For both models, we use $C_m=1 \mu$F/cm$^2$ while the
diffusion constant is chosen to be
$D=$0.0005 cm$^2$/ms for the FK model and
$D=$0.001 cm$^2$/ms for the LR model.
Simulations are carried out with a discretization of 0.025 cm, using a 
5-point stencil, and a time step of 0.025 ms, using explicit Euler integration. 
For both models,  the conduction velocity along a cable  is within the 
electophysiological range: 51 cm/s for the FK model  and 33 cm/s for the LR model. Errors in
direct simulation results are reported as standard deviations.

\section{Results using direct simulations}

\begin{figure}[ht]
	\centering
	\includegraphics[width = 0.7\linewidth]{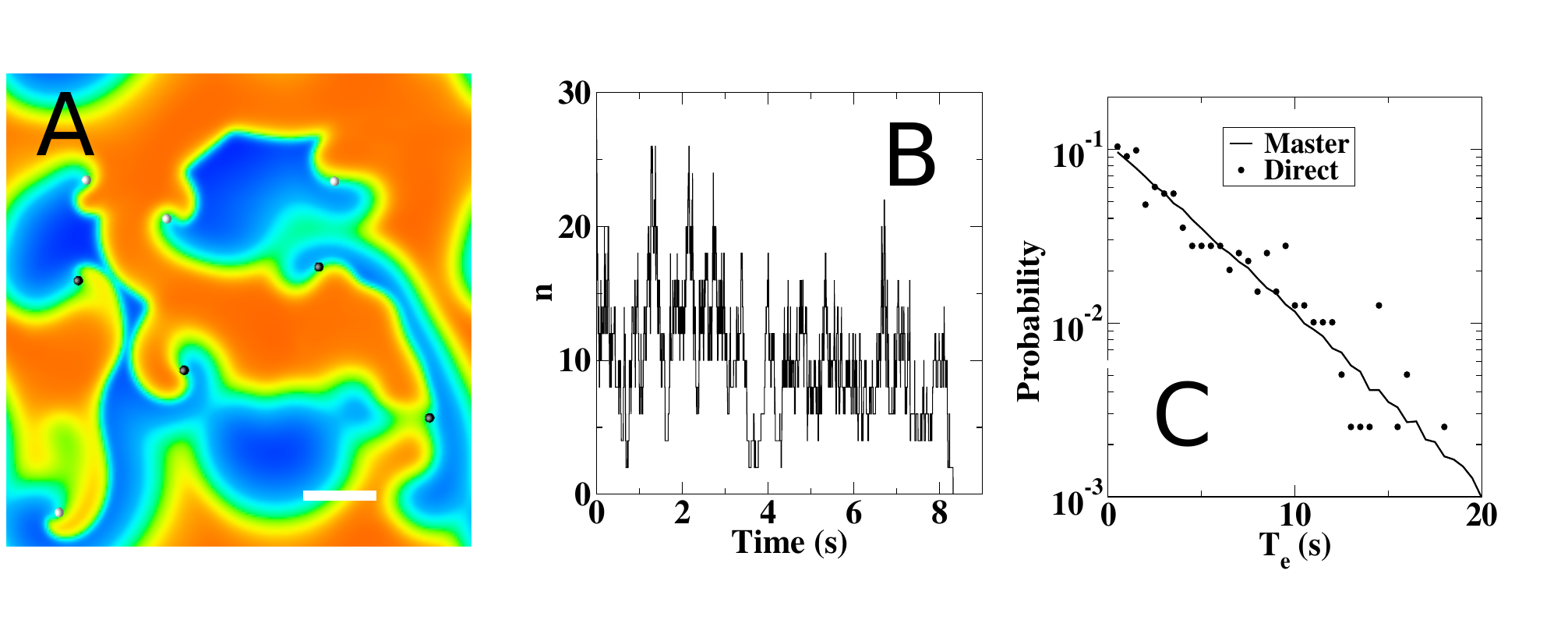}
	\caption{{\bf Direct numerical simulations provide statistics of spiral tip dynamics.}
		{\bf A:} Snapshot of a simulation of the LR model in a 7.5$\times$7.5 cm computational domain with periodic boundary conditions. 
		The voltage is represented using a color code with red (blue) corresponding to depolarized (repolarized) tissue. The location of the tips of counter- and clockwise rotation spiral waves are shown in black and white, respectively. (scale bar: 1cm) {\bf B:} Typical time trace of the number of spiral tip pairs. For this particular simulation, spiral tips spontaneously extinguished after 8.3 s.
		{\bf C:} Distribution of termination times for the direct simulations (symbols, computed using 400 termination events) and the master equation (solid line, computed using 10000 termination events).} 
	\label{fig_snap}
\end{figure}

Starting with a random initial condition that contains multiple spiral waves, we solve the reaction-diffusion  equation and keep track of the number of spiral tips using a standard algorithm  (Fig. \ref{fig_snap}A) \cite{FenKar98}. 
The number of tips fluctuates and the simulation ends after time $T_e$ when the number of spiral tips reaches 0 (Fig. \ref{fig_snap}B). 
We can compute the distribution of these termination times 
by repeating the simulations many times,  starting with different and independent initial conditions.
These conditions are created by perturbing a multi-spiral state with randomly placed stimuli in the form of a current stimulus of duration 20 ms and strength several times the excitation threshold. After perturbation, the system is allowed to evolve for another 100 ms before measurements are started. 
   Our simulations reveal that this distribution is exponentially distributed, indicating that  
   spiral wave termination can be well described as a Poisson process (Fig. \ref{fig_snap}C). 

\begin{figure}[ht]
	\centering
	\includegraphics[width = 0.5\linewidth]{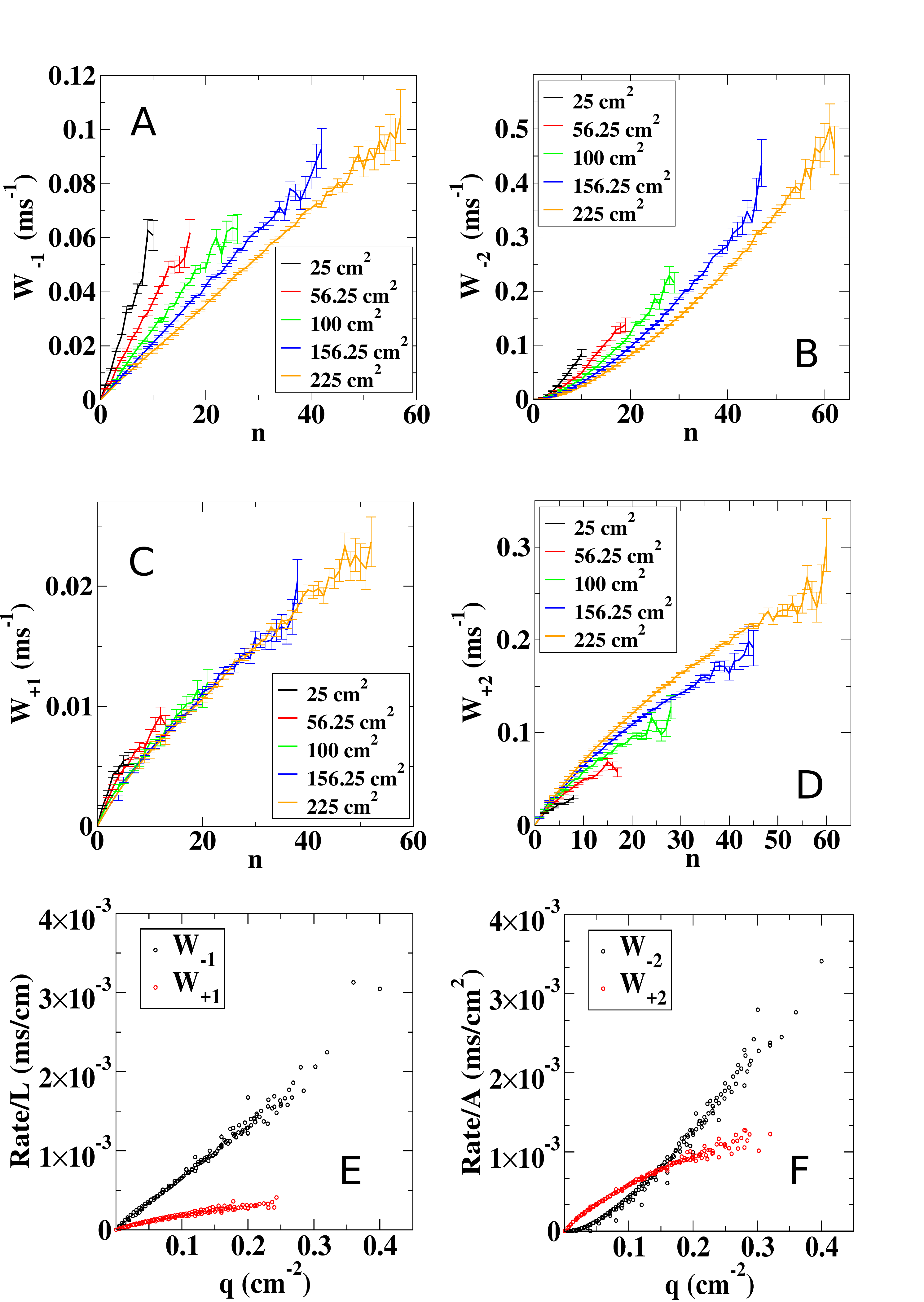}
	\caption{{\bf Transition rates computed using direct simulations.} {\bf A-D:}The birth and death rates for  
		$n\rightarrow n-1$ (A),  $n\rightarrow n- 2$ (B),
		$n\rightarrow n+1$ (C),  $n\rightarrow n+ 2$ (D) 
		computed in a square geometry of various sizes with 
		non-conducting boundaries. Error bars represent standard deviation.
		{\bf E:} The $W_{\pm1}$ rates, normalized by the perimeter of the domain, as a function of the density of tips, $q=n/A$. {\bf F:} The $W_{\pm2}$ rates, normalized by the area of the domain, as a function of the density of tips.}
	\label{fig_rates}
\end{figure}

Next, we compute the birth and death rates 
as a function of the number of tips $n$ using different domain sizes
with non-conducting boundaries 
 by quantifying the number of 
transitions per time interval (Fig. \ref{fig_rates}A-D).
For fixed computational time and domain size, the number of transitions will depend on the 
number of tips. It will  become smaller as $n$ increases and 
no transitions will be recorded above some critical value of $n$. 
Here, to increase accuracy, 
 we only consider rates that are computed using 
at least 100 transitions in the simulation.
As a consequence, rates are  computed up to a 
certain maximum value 
of $n$.
In addition, for increasing domain sizes, transitions  for 
small  $n$ become increasingly rare. As a result, in large domains, the number of 
recorded transitions for small values of $n$ 
 may not reach  100. The rates corresponding to these values 
 of $n$ are therefore   not included.
 
Examining the computed rates, we see that $W_{-1}$ depends linearly on the number of spiral tips for all domain sizes (Fig. \ref{fig_rates}A). The remaining rates, however, show a more complex dependence on the 
number of tips, indicating
the existence of non-trivial long-range interactions between spiral tips
(Fig. \ref{fig_rates}B-D).  
As a result, the rate curves are not easily fit  by simple rational 
functions. 
Therefore, 
we employ a smoothing spline fit to the data to determine  rates corresponding to transition events 
with less than the minimum number. 

We can also compute the $W_{\pm2}$ rates for domains that 
contain  periodic boundary conditions. The results of these simulations 
are  shown in the Supplemental Material \cite{supplemen} and 
are qualitatively similar to the results presented in Fig. 2.
As a consistency check, we can use these rates to compute the distribution of termination times.
As expected, this distribution is exponential and
agrees well with the one computed using direct simulations
(Fig. \ref{fig_snap}C).
In addition, we compute the rates for the 
FK model and show the results in   the Supplemental Material \cite{supplemen}.

Importantly, we find that at large $A$ all rates collapse onto a single curve 
when plotted as a function of the density $q=n/A$.  Specifically, 
the $W_{\pm 2}$ rates are found to scale with the 
area as $W_{\pm 2} (n) \sim A w_{\pm 2} (q) $ 
(Fig. \ref{fig_rates}F), 
 indicating that the birth and death rates only depend on the 
 density and that tips are well-mixed.
 Furthermore, 
the $W_{\pm 1}$ rates scale with the perimeter $L$ 
 as $W_{\pm 1} (n) \sim L w_{\pm 1} (q) $ (Fig. \ref{fig_rates}E).
 Here, and in the following, we will take the continuum limit such 
 that $q$ and functions that depend on this variable are considered 
 to be continuous. 
Note that this observed linear scaling of $W_{-1}$ with $L$  implies that the 
death rate is proportional with the length of the non-conducting boundary and that creating ablation lesions will increase this rate.
Furthermore, such scaling is expected if single tips annihilate through simple collision processes and get created near the boundaries. 
Similar 
scaling behavior is found 
 for periodic boundary conditions and for the FK model,
 as shown in the Supplemental Material \cite{supplemen}. 

\begin{figure}[ht]
	\centering
	\includegraphics[width = 0.5\linewidth]{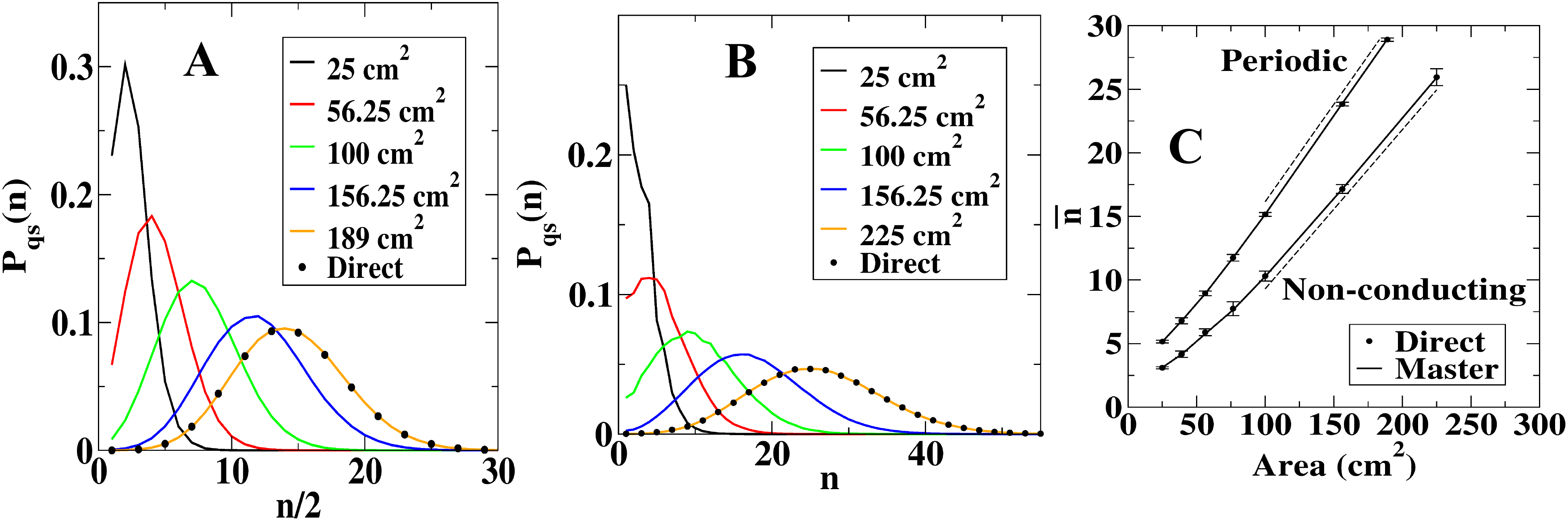}
	\caption{{\bf Dependence of spiral tip dynamics on the domain size.}
		{\bf A, B:} The quasi-stationary distribution for periodic (A) and non-conducting boundary conditions (B) using different domain sizes as computed using the transition matrix. The symbols show the quasi-stationary distribution as computed using the direct simulations. {\bf C:} The average number of tips as a function of the area of the computational domain, computed using direct simulations (symbols) and using the master equation approach (line). The dashed curves are straight lines.   }
	\label{fig_area}
\end{figure}

\section{Results using transition rates}

  Once the transition rates are determined, it is straightforward to 
 compute the 
quasi-stationary distribution $P_{qs}(n)$ using the transition matrix 
at minimal computational cost  (Fig. \ref{fig_area}A, B) \cite{newman2010networks}. 
For small domains, this can be carried out using the rates obtained in 
the simulations while for  larger domains,
where the rates for small $n$ cannot be computed accurately, 
we can use the interpolated rates.
As the domain size increases, the distribution shifts to larger values of $n$, and becomes more symmetric around its peak. 
The average  number of tips, $\bar{n}$, increases with system size and our simulations reveal that  
it depends linearly on the area of the computational domain for both boundary conditions (Fig. \ref{fig_area}C). Results for the FK model and for 
periodic boundary conditions are qualitatively similar (see the Supplemental Material \cite{supplemen}).

For geometries that do not contain any non-conducting boundaries it is 
possible to derive closed-form solutions for the quasi-stationary distribution. 
In this case, $n$ is always even and tips will be created and annihilated in pairs 
such that $W_{\pm 1}=0$.
The quasi-stationary distribution can be obtained by setting the left hand side   in Eq. \ref{eq_master} to zero, resulting in the recursion relationship
\begin{eqnarray}
P_{qs}(n) = P_{qs}(0) \prod_{j=2}^{n} W_{+2}(2j-2)/W_{-2}(2j) 
\end{eqnarray}  
where $P_{qs}(0) $ can be determined by the normalization condition $\sum_{n=0}^{\infty}P_{qs}(n)=1$ \cite{Gard_book09}.
The average number of tips can be obtained using 
\begin{equation}
\frac{d<n>}{dt}=\frac{d}{dt}\left( \sum_{n'=1}^{\infty}n'P(n')\right )
\end{equation}
which results in
a deterministic equation \cite{Gard_book09}
\begin{equation}
\frac{dn}{dt}=2W_{+2}(n)-2W_{-2}(n)
\end{equation}
and 
a  deterministic stationary state determined by
$W_{+2}(n^*)=W_{-2}(n^*)$. 
The maximum value of the quasi-stationary
distribution occurs for $P_{qs}(n-2)/P_{qs}(n)\approx 1$, 
corresponding to 
$W_{+2}(\bar{n}-2)=W_{-2}(\bar{n})$. Therefore, for large values of $A$ the
stochastic average number can be well approximated by
deterministic average number, $ n^* \approx \bar{n} $.
Furthermore, using our numerically found scaling, we obtain
$w_{+2}(q^*)=w_{-2}(q^*)$, where $q^*=n^*/A$. Hence,  the average density
is independent of the area and $n^*$, and  $\bar{n}$ scales with $A$,
consistent with the scaling found in the simulations (Fig. \ref{fig_area}C).
For domains that contain non-conducting boundaries, the $\pm$1 rates are no longer 
zero and 
the corresponding
deterministic equation reads
\begin{equation}
\frac{dn}{dt}=2W_{+2}(n)+W_{+1}(n)-2W_{-2}(n)-W_{-1}(n).
\end{equation} 
Using our obtained scaling, we have 
\begin{equation}  2w_{+2}(q^*)+w_{+1}(q^*)/\sqrt{A}-2w_{-2}(q^*)-w_{-1}(q^*)/\sqrt{A}=0
\end{equation} 
and 
for large areas, the average number of tips will again scale
linearly with the area.

  \section{Mean episode duration}

\begin{figure}[ht]
	\centering
	\includegraphics[width = 0.7\linewidth]{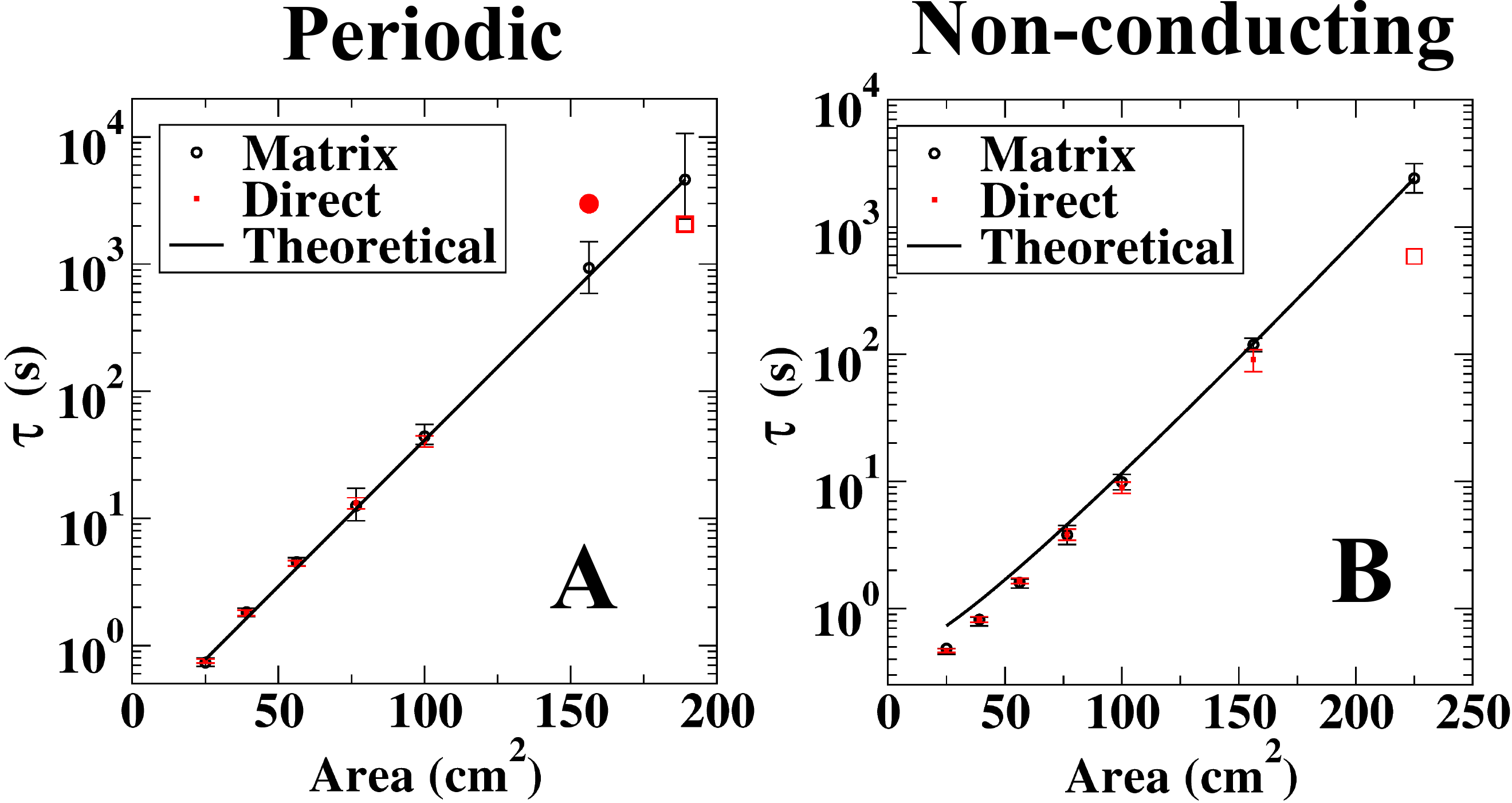}
	\caption{{\bf Termination times as a function of the system size.} {\bf A, B:} $\tau$ as a function of the area of the computational domain from direct simulations using domains with periodic boundary conditions (A) and non-conducting boundaries (B) (red symbols). Also shown are the results  from the master equation approach (black symbols) and from the closed-form expression obtained using the WKB analysis (solid line). The red square represents the result of a single termination event computed using direct simulations while the solid red circle represents the computed time for a single computation that did not result in a termination.  }
	\label{fig_wkb}
\end{figure}

To find the mean episode duration  $\tau$ in the direct simulations, 
we  average 
the termination times $T_e$ obtained from each independent simulation.
This computation becomes more and more time consuming as $A$ increases 
since termination becomes less and less likely.
As a consequence, the number of determined termination events we consider vary  from 400 for small domains to less than 
10 for the largest areas still amenable to direct simulations.  
Our results reveal that  
$\tau$ displays an exponential dependence on the size of the domain, 
both for periodic  
(red symbols, Fig. \ref{fig_wkb}A) and non-conducting 
boundary conditions (red symbols, Fig. \ref{fig_wkb}B), 
consistent with earlier studies \cite{Qu06}. 

Rather than using direct simulations to determine an average value for 
$T_e$, it is straightforward to use the 
interpolated transitions rates and the resulting 
transition matrix to compute $\tau$ using 
simple matrix operations \cite{newman2010networks}.
For this, we construct a transition matrix $Q$ for all transient states $n>0$, with elements $Q_{ij}$ representing the probability of transitioning from state $i$ to state $j$.  The probability of reaching state $j$ from state $i$ in $t$ steps is then given by the $ij^{th}$ entry of $Q^t$.  Summing this over all time results in the so-called fundamental matrix $N = I + Q + Q^2 + ... = (I-Q)^{-1}$, where $I$ is the identity matrix.  Each element of the fundamental matrix $N_{ij}$ represents the mean duration our system will spend in state $j$ given an initial state $i$, which can be used to determine the quasi-stationary distribution.  
Moreover the mean time to extinction $\tau$ is given by $N\vec{e}$, where $\vec{e}$ is a column vector of ones.
The 
confidence intervals for $\tau$ are computed through bootstrapping as follows.  First we resample each transition rate by drawing a value from a binomial distribution with probability equal to the original transition rate and 
using the number of recorded transitions from the direct simulation.  We then proceed by interpolating these resampled transition rates and computing $\tau$ from these interpolated transition rates.  This is computed for 1000 trials and the confidence interval is determined from the 5th to the 95th percentile of the resulting $\tau$ across all trials \cite{web}.

\begin{figure}[ht]
  	\centering
  	\includegraphics[width =0.5\linewidth]{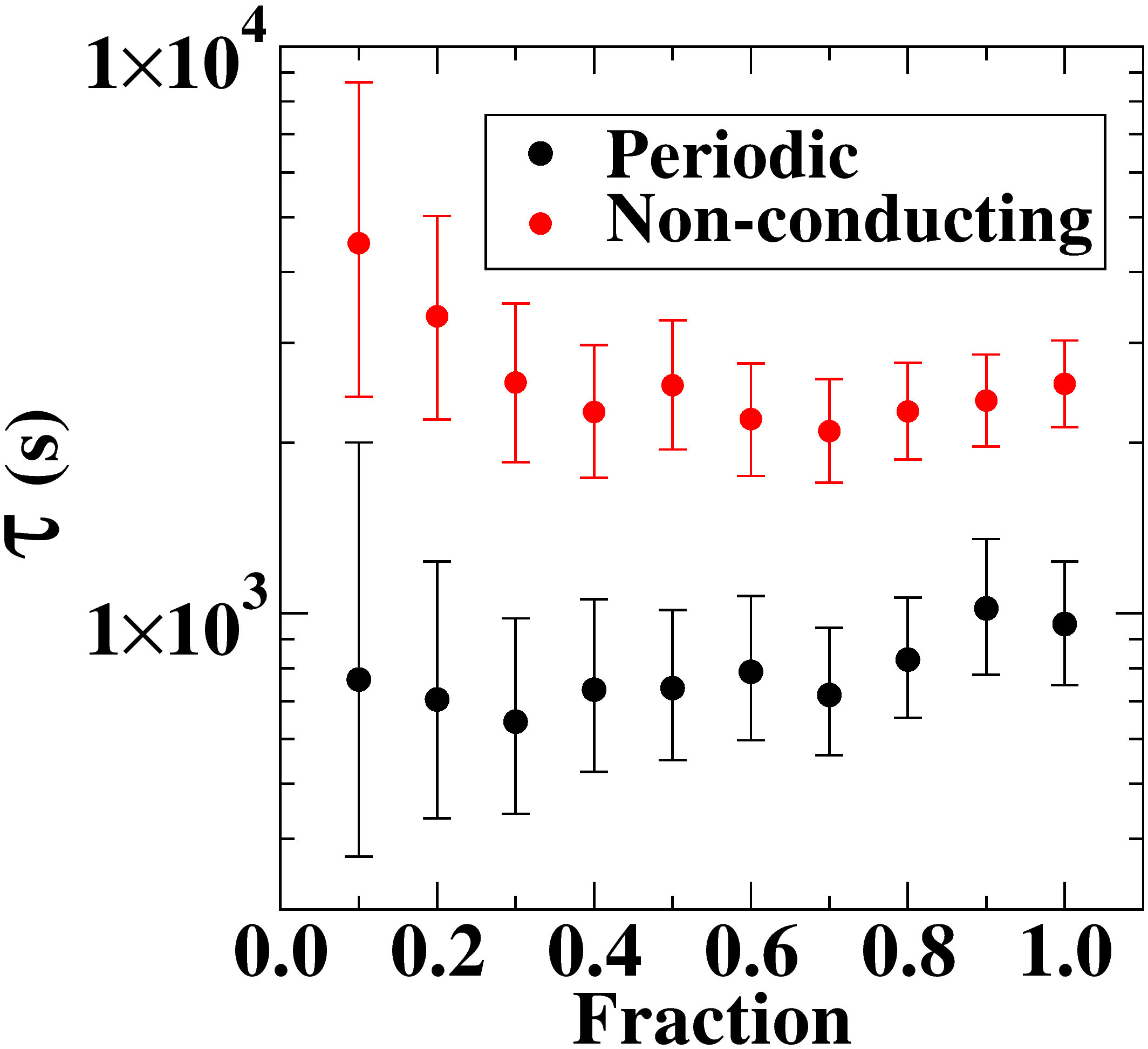}
  	\caption{{\bf Accuracy of the master equation approach.} The mean episode duration computed using the master equation as a function of the fraction of computational data obtained using direct simulations.
  		Data segments of size   indicated by fraction were started at random positions.
  		Shown are the results  for $A=225 cm^2$ for the non-conducting case and for $A=189.0625 cm^2$ for the periodic case.
  		The error bars are determined using bootstrapping and represent the 5\% and 95\% confidence interval.
  	}
  	\label{fraction}
  \end{figure}

The resulting values for $\tau$ agree well with the direct numerical simulations (black symbols, Fig. \ref{fig_wkb}A and B).
Importantly,  using the transition matrix allows us to estimate the mean episode duration for system sizes where determination of mean episode duration with direct simulations  is impossible.
For example, directly simulating a single extinction event on a domain with area $A=225 \mu m^2$ and non-conducting boundaries was found to take approximately 100 hours of CPU time. Estimating $\tau$ from this single event is not useful as the error is large and generating a sufficient amount of termination events is not practical. Furthermore, for other larger domain sizes our direct simulations failed 
to produce  a single termination event, even after 7 days of CPU time. 
Using the 
interpolated
transition rates computed from this single, non-terminating event, however, we are still 
able to use the transition matrix (Fig. \ref{fig_wkb}A and B) to predict the 
mean episode duration.  Moreover, $\tau$ can already  be estimated using only 
a fraction of the data, and thus simulation time, further demonstrating the power of the  approach. This is shown in Fig. 
\ref{fraction} where we plot $\tau$ as a function of the fraction of computational data from a direct simulation.
Obviously, for larger fractions, the errors in the transition rates become 
smaller, resulting in smaller confidence intervals. 
 Furthermore, the
 mean termination time converges as the fraction increases and can be 
 reasonably well estimated from a small fraction of the entire dataset.

\section{Scaling results}

\begin{figure}[ht]
	\centering
	\includegraphics[width=0.5\linewidth]{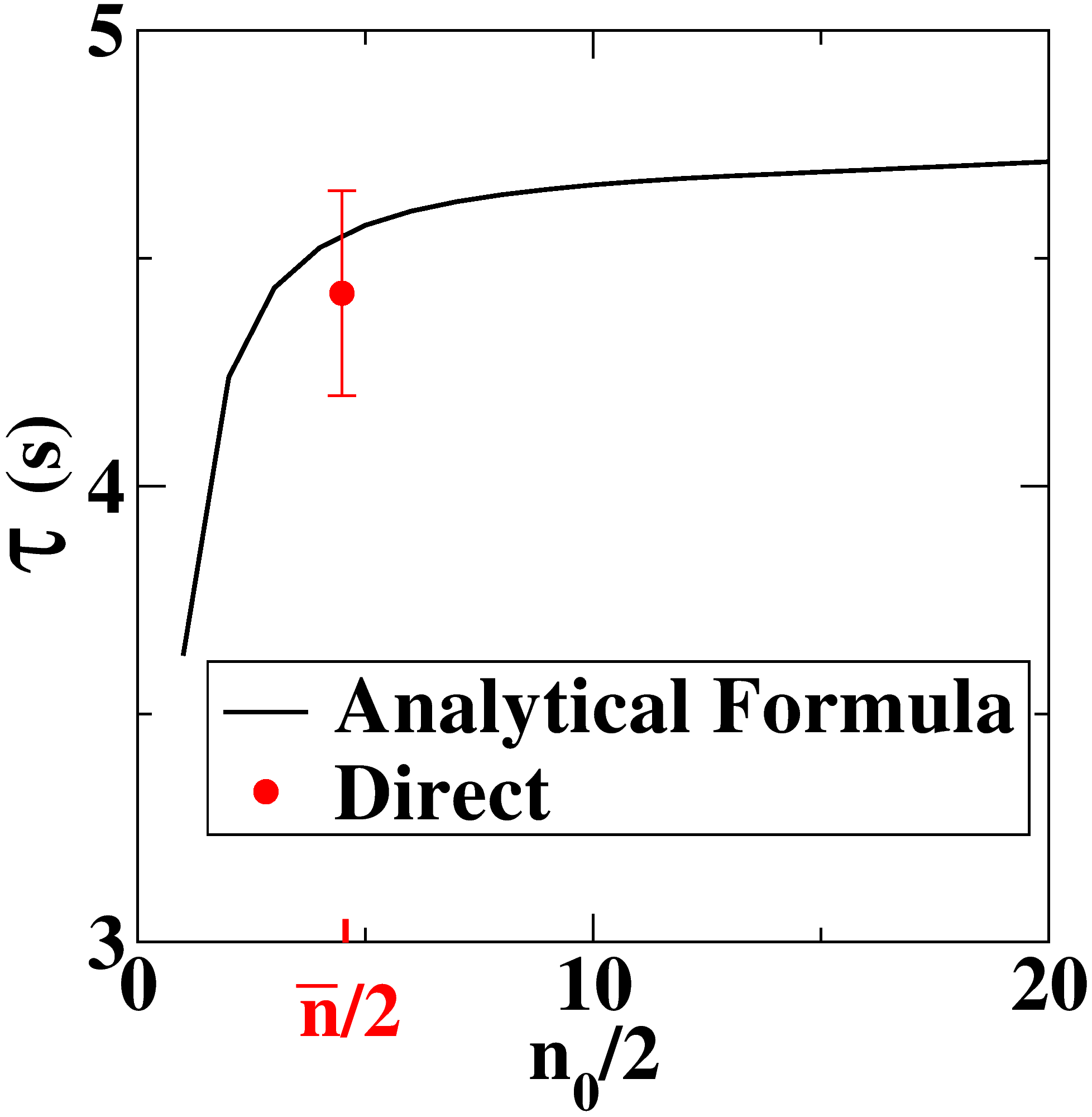}
	\caption{{\bf Analytical formula for periodic boundary conditions.} Mean episode duration $\tau$ as computed using the analytical formula  in the main text as a function of the initial number of tips, $n_0$. Symbol corresponds to the result from direct simulations ($\bar{n}\approx 9$). Shown are the results for the LR, using a domain of size 7.5cmx7.5cm. }
	\label{analytic}
\end{figure}

We can also use our stochastic analysis of termination to  determine the scaling of $\tau$ with the area.
For periodic boundary conditions, it is  possible to obtain 
an analytical expression for  $\tau$ \, \cite{Gard_book09}:
\begin{eqnarray}
\tau(n_0)=\sum_{k=1}^{n_0/2}\phi(2(k-1))\sum_{j=k}^{\infty}\frac{1}{\phi(2j){W}_{+2}(2j)} 
\label{eq_mtt_per}
\end{eqnarray}
where $n_0$  is the initial number of spiral tips,  $\phi(k)=\prod_{i=1}^{k/2}{W}_{-2}(2i)/{W}_{+2}(2i)$, and $\phi(0)\equiv 1$.
Using the numerically determined rates we find that 
$\tau$ quickly converges as $n_0$ becomes large and that  $\tau(\bar{n})$ agrees well with the 
values obtained using the numerical methods. 
This is shown explicitly in Fig. \ref{analytic} which 
plots the mean episode duration as a function of the 
initial number of tips. 

To determine the scaling with the area we  focus on the
first term of this expression, $\tau(2)=
\sum_{j=1}^{\infty}(\phi(2j){W}_{+2}(2j))^{-1}$. We can write $\phi(2j)$ as
\begin{equation}
{\rm ln} \left[ \phi(2j) \right]=\sum_{z=1}^{j} {\rm ln} \left[ \frac{W_{-2}(2z)}{W_{+2}(2z)}
\right] \approx -\frac{A}{2} \int_{2/A}^{x} {\rm ln} \frac{w_{+2}(s)}{w_{-2}(s)} \ ds
\label{}
\end{equation}
where we have used the fact that  the transition rates scale with the
area $A$ and have defined   $s=2z/A$ and  $x=2j/A$.
As a result, the mean episode duration becomes
\begin{equation}
\tau \approx \int_{0}^{\infty} \frac{ {\rm exp}\left[ A \int_{2/A}^{x} {\rm ln} \sqrt{\frac{w_{+2}(s)}{w_{-2}(s)}}      
	ds \right]}{2w_{+2}(x)} dx.
\label{}
\end{equation}
For large $A$, this integral will be sharply peaked around $q^*$.
Thus, $\tau$ has the following scaling behavior \cite{krapivsky2010kinetic}:
\begin{equation}
\tau \sim {\rm exp} \left[ A \int_{2/A}^{q^*} {\rm ln} \sqrt{ \frac{w_{+2}(s)}{w_{-2}(s)}}
ds \right]
\label{scaling}
\end{equation}
and, as an immediate consequence of the observed scaling of 
the transition rates, we  
find that  $\tau$ scales exponentially with the area, consistent with 
our direct numerical results (Fig. \ref{fig_wkb}A). 

We can also use approximation methods to determine the scaling of the 
mean episode duration by viewing the number of spiral tips as a stochastic population in a  metastable state.
This approach is particularly useful 
for domains containing non-conducting boundaries, for which 
it is no longer possible 
to derive an exact expression for $\tau$. 
 As long as $A$, equivalent to the total population size in models of population biology, 
is sufficiently large, we can use
a WKB approximation. 
In this approximation, the quasi-stationary distribution is assumed to obey $P_{qs}(q) \sim e^{-A S(q)}$ where $S(q)$ is a function  
called the action. 
We can now use our obtained scaling $W_r(n) = A^{r/2} w_r(q)$, together with  the assumed form of $P_{qs}(q)$, and substitute them into the 
stationary form of Eq. 1. This equation is written in terms of the 
continuous rescaled variable $q=n/A$ so that 
$n -r \rightarrow q - r/A$ \cite{assaf2010extinction}.     
For the periodic case, we take $S(q)=S_0(q)+O(A^{-1})$ while 
for absorbing boundaries, since the scaling of the $\pm 1$ rates goes as $\sqrt{A}$ while the $\pm 2$ rates go as $A$, we use $S(q)=S_0(q)+A^{-1/2}S_1(q) + O(A^{-1})$.    The resulting 
equation can then be expanded in terms of $1/A$
which  yields, to $O(1)$, a Hamilton-Jacobi equation
\begin{equation}
H(q,p) = \sum_r A^{r/2} w_r(q) (e^{rp}-1) = 0
\end{equation}
where $p(q)=\partial S/\partial q$ is the fluctuation momentum \cite{dykman1994large,assaf2010extinction}.

From the Hamiltonian $H$ we can define the dynamics of $p$ and $q$ 
using $\frac{dp}{dt}=-\frac{\partial {H}}{\partial q }$ and
$\frac{dq}{dt}=\frac{\partial {H}}{\partial p }$. 
The  non-trivial solution of   $H(q,p_a(q))=0$ corresponds to the 
activation trajectory in the $q,p$ phase space \cite{assaf2010extinction, kubo1973fluctuation,dykman1994large,kessler2007extinction,doering2005extinction}. This trajectory  
 describes the most probable path along which the system evolves from 
 the metastable state $(q^*,0)$ to a point $q$ in phase space. 
 Since we are interested in extinction,
 we will consider the trajectory that connects $(q*,0)$  with 
 $[0,p(0)]$, the so-called ``optimal'' path to extinction \cite{dykman1994large}.
 This $q,p$ phase space, along with the activation trajectory, is shown in Fig. \ref{phase_space} for periodic boundaries
 and in the 
 Supplemental Material for the FK model \cite{supplemen}.  
 The optimal path  can be determined  numerically but 
 can also be determined using
 approximate closed-form relations.
 Specifically, 
 for periodic boundary conditions, we find
 \begin{equation}
 S_0 = \int_{q^*}^{2/A}\ln{\gamma_0} \ dq,
 \end{equation}
 and $\gamma_0 = \sqrt{\frac{w_{-2}}{w_{+2}}}$. 
 In Fig. \ref{phase_space}, this corresponds to 
 the area between the activation trajectory and the $q$ axis, represented by
 the shaded part. 
 Thus, we find that the mean episode duration scales as 
 $\tau \sim e^{A S_0}$,
 consistent with Eq. \ref{scaling}.
  For absorbing boundaries, we can  solve for $S_1$ perturbatively, yielding
 \begin{equation}
 S_1 = \int_{q^*}^{2/A}\frac{(\gamma_0 w_{+1} - w_{-1})(\gamma_0-1)}{2\gamma_0(w_{+2}-w_{-2})}dq.
 \end{equation}

 \begin{figure}[ht]
 	\centering
 	\includegraphics[width = 0.5\linewidth]{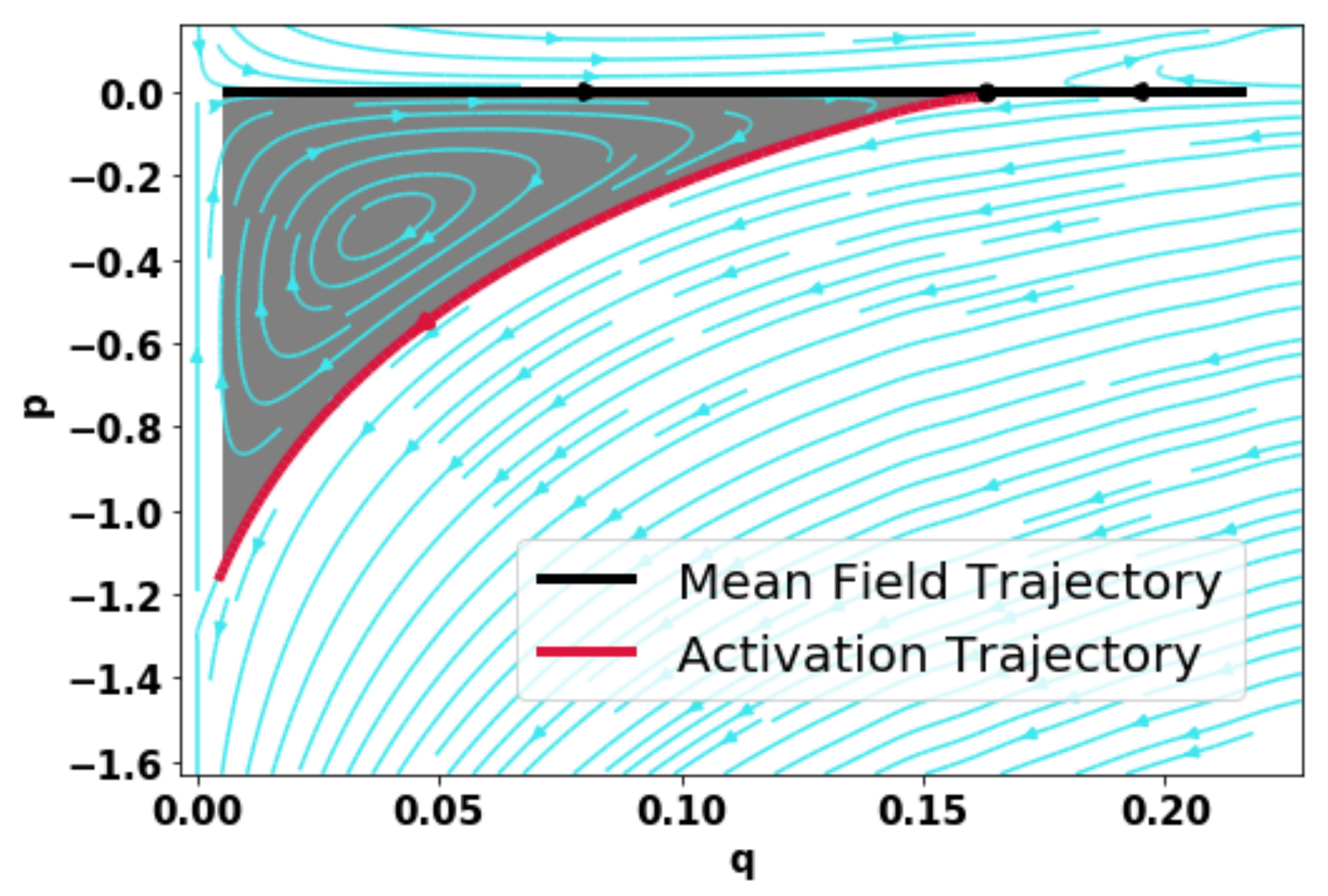}
 	\caption{{\bf WKB approach to spiral tip dynamics.} Phase portrait  of the Hamiltonian dynamics in $q$, $p$ space for periodic boundary conditions, showing the 
 		activation trajectory of the WKB Hamiltonian (red line). The shaded area represents the exponential factor $S_0$. }
 	\label{phase_space}
 \end{figure}

\begin{figure}[ht]
  	\centering
  	\includegraphics[width =0.5\linewidth]{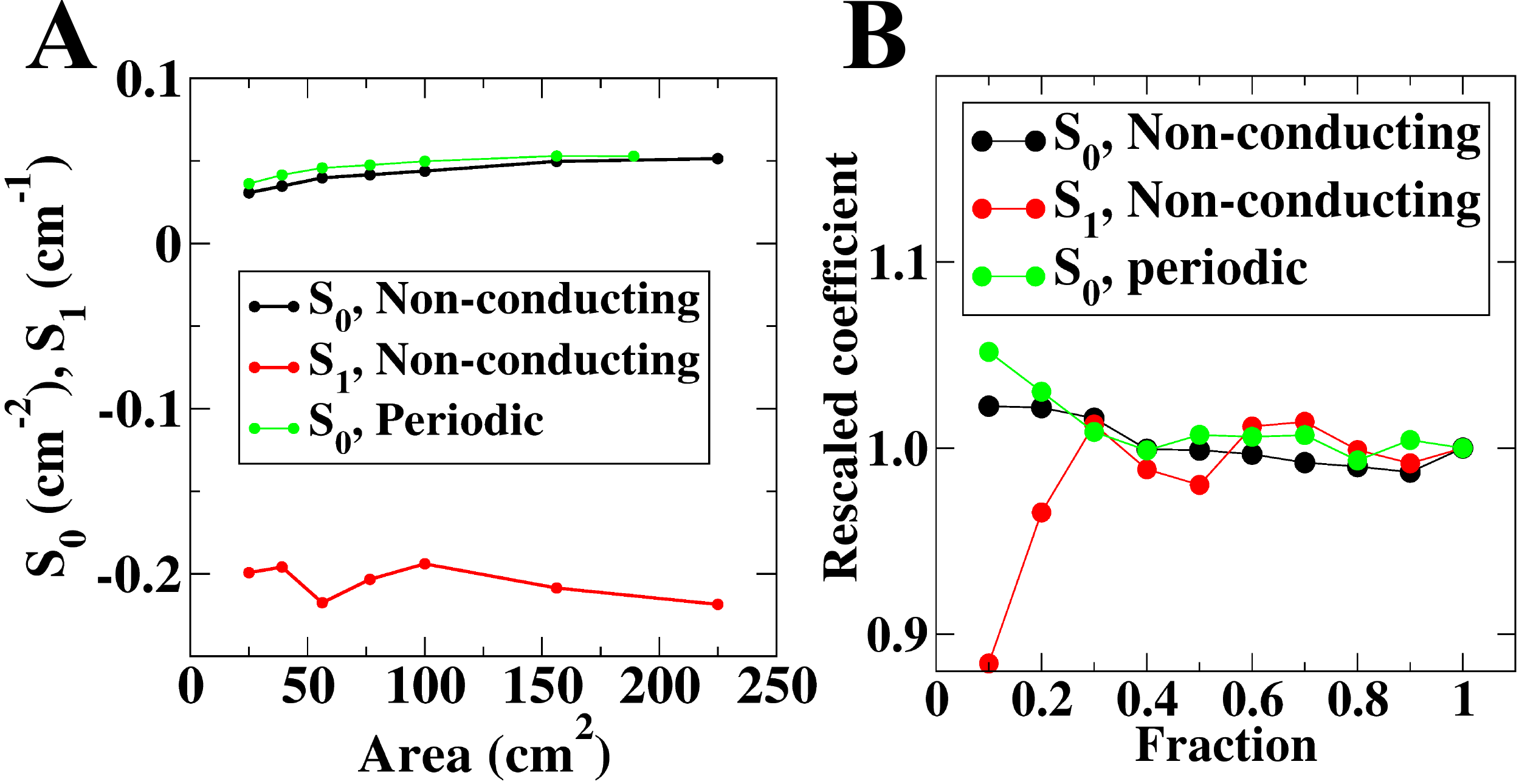}
  	\caption{{\bf WKB parameters as a function of area size for the LR model} A: The exponential coefficients $S_0$ and $S_1$ as a function of domain size for periodic and non-conducting boundaries.
  		B: The exponential factors $S_0$ and $S_1$, rescaled by their value computed at the largest computational data set, as a function of the fraction of computational data obtained using direct simulations.
  		Data segments of size   indicated by fraction were started at random positions.   Shown are the results  for $A=225 cm^2$ for the non-conducting case and for $A=189.0625 cm^2$ for the periodic case.
  	}
  	\label{converge}
  \end{figure}

Using our formulation, we can now compute the mean episode duration for 
any system size once the rates for a single domain $\hat{A}$ are 
determined. Specifically, after computing for this particular domain both 
$\hat{S}_0$ and $\hat{S}_1$ and  the corresponding mean episode duration
$\hat{\tau}$, 
we can find $\tau$ as a function of the system size using:
\begin{eqnarray}
\tau(A) \approx \hat{\tau} e^{ (A-\hat{A}) \hat{S}_0 + (\sqrt{A}-\sqrt{\hat{A}})\hat{S}_1}.
\label{tau_eqn}
\end{eqnarray}    
This scaling law for $\tau$  agrees well with the  values of $\tau$ 
computed from the master equation, especially for larger values of $A$  (Fig. \ref{fig_wkb}A, B), 
justifying the WKB approximation.
For these 
larger domain sizes, the   factors $S_0$
and $S_1$ converge, making 
the estimate from Eq. \ref{tau_eqn}  to be more accurate, 
as shown in Fig. \ref{converge}A.
 This is consistent with the 
obtained quasi-stationary distributions which become 
more symmetric around their peak value for larger domain size (Fig. 3A and B), rendering the WKB approximation more accurate. Furthermore, as is the case for $\tau$  (Fig. \ref{fraction}), both 
 $S_0$ and $S_1$ can be estimated using 
 the interpolated rates and
  only a fraction of the direct simulation data (Fig. \ref{converge}B). Thus, accurate estimates for arbitrary domain sizes 
do not require simulating actual termination events.
Finally, the exponential scaling of $\tau$ with 
system size $A$   reveals that, even though spiral wave driven fibrillation will always  terminate,
its mean episode duration depends critically on the size of the heart. 
Of course, this result is valid as long as the specifics of the model do not change. Other factors, including changes in electrophysiological parameters,
can have an effect of mean termination duration. 
For large values of $A$, $\tau$ can be large while   
for very small values of $A$ as found, for example, in rodents, the
mean episode duration will be well below 1 s.
These findings provide a mechanistic underpinning for the well-established
critical mass hypothesis which posits that fibrillation only occurs in hearts of a minimum size \cite{garrey1914nature,Qu06, Byretal05}. 
  
 \section{Summary}
   
In summary, we present a novel approach to quantify spiral wave dynamics in spatially extended domains. This approach recasts the problem into a master equation, after which statistical physics methods can be employed. 
Our approach is valid for any model exhibiting SDC, including electrophysiological models, and any geometry.
Key in this approach are the transition rates, which can be computed 
numerically from a limited set of direct simulations.
Future work can include applying our approach to geometric models of 
atria. Geometry data are routinely obtained in patients and, combined
with our analysis, might result in more patient-specific approaches for 
AF.  In addition, it would be interesting to further study
the dependence of the rates on the number of tips. If, for example, 
rational functions 
for these rates can be derived, it will be possible to obtain 
analytical expressions for $\tau$.  
Finally, it would be interesting to compare  scaling  
 of atrial fibrillation in healthy and diseased atria by 
 simulating appropriate electrophysiological models.  
Thus, stochastic analysis of spiral wave reentry has the potential to be an important step towards determining optimal therapeutic interventions aimed at minimizing the duration of AF episodes.

\section{Acknowledgements}
We thank Brian A. Camley and David A. Kessler for valuable suggestions.
We gratefully acknowledge support from  the National Institutes of Health (R01 HL122384) and the American Heart Association (16PRE30930015).

\end{document}